\documentclass[aps]{revtex4}
\usepackage{graphicx}
\usepackage{color}
\RequirePackage{amsfonts,amssymb,amsmath}
\usepackage[normalem]{ulem}
\usepackage{hyperref}

\begin{document}
\title{ Modeling the quantum evolution of the universe through classical matter}
\author{Jo\~ao Paulo M. Pitelli} 
\email{joao.pitelli@ufabc.edu.br}
\affiliation{Centro de Matem\'atica, Computa\c{c}\~ao e Cogni\c{c}\~ao, UFABC, 09210-70, Santo Andr\'e, SP, Brazil}
\author{Patricio S. Letelier} 
\email[]{Deceased.}
\affiliation{Departamento de Matem\'atica Aplicada-IMECC,
Universidade Estadual de Campinas,
13081-970 Campinas,  S.P., Brazil}


\begin{abstract}

It is well known that the canonical quantization of the Friedmann-Lema\^itre-Robertson-Walker (FLRW) filled with a perfect fluid leads to nonsingular universes which, for later times, behave as their classical counterpart. This means that the expectation value of the scale factor $\left<a\right>(t)$ never vanishes and, as $t\to\infty$, we recover the classical expression for the scale factor. In this paper, we show that such universes can be reproduced by classical cosmology given that the universe is filled with an exotic matter. In the case of a perfect fluid, we find an implicit equation of state (EoS). We then show that this single fluid with an implict EoS is equivalent to two non-interacting fluids, one of them representing stiff matter with negative energy density. In the case of two non-interacting scalar fields, one of them of the phantom type, we find their potential energy.  In both cases we find that quantum mechanics changes completely the configuration of matter for small values of time, by adding a fluid or a scalar field with negative energy density. As time passes, the density of negative energy decreases and we recover the ordinary content of the classical universe. The more the initial wave function of the universe is concentrated around the classical big bang singularity, the more it is necessary to add negative energy, since this type of energy will be responsible for the removal of the classical singularity.
\end{abstract}
\maketitle

\section{Introduction}
\label{Introduction}
Classical singularities are always present in cosmological models, provided that the matter content of the universe fills some reasonable requirements, called energy conditions, which basically state that gravity is attractive \cite{ellis}. This is a burden of general relativity and there is a general agreement that a full quantum theory of gravitation will solve this problem, either indicating how to deal with the singularities or excluding them at all. Besides quantum gravity, there are many ways to circumvent the initial big bang sigularity in cosmological models, even at the classical level. For example, there are $f(R)$ theories of gravity which avoids the singularity in FRW models (the conditions for a bounce in $f(R)$ theories where sudied in \cite{carloni}). Theories with a scalar field in the presence of a potential can also be free of singularities (see, for instance, \cite{alt}-\cite{kamenshchik} for a classical treatment and \cite{kamenshchik}-\cite{ashtekar} for the quantum mechanical point of view.).

The first bouncing models where obtained by Novello and Salim \cite{novello}, and Melnikov and Orlov \cite{melnikov} in the 70's, but they were taken more seriously in the late 90's, when the discovery of the accelerating expansion of the universe made the idea of violation of the energy conditions more reasonable. For a review about the models and the importance of bouncing cosmologies, see \cite{novello2}.

 There are several pieces of evidences suggesting that quantum mechanics will be able to exclude the classical singularities on cosmological models. It was shown in \cite{lemos,alvarenga} that the quantum evolution of the FLRW universe filled with a perfect fluid with EoS $p=w\rho$ is non-singular and matches the classical evolution when $t\to\infty$. In this context, two works showed that the energy conditions for matter can be quantum mechanically violated, a necessary requirement for the exclusion of singularities. In the first one \cite{pinto-netoref}, through the quantization of the FRW universe in the presence of two fluids, dust plus radiation, Pinto-Neto et al. show that, in this model,  in the far past the universe was filled with a classical exotic dust (dust with negative density) and as the universe expands the conventional dust and radiation prevails. In the second one \cite{letelier1}, the authors quantize the Riemann tensor in the tetrad basis and, through the Eintein equation, find expressions for the quantized fluid, showing that it violates the strong energy condition in the quantum domain, recovering the conventional behavior at the classical regime. In Ref. \cite{bojowald}, Bojowald showed that, in Loop Quantum Cosmology the big-bang singularity is also removed, by analysing the inverse scale factor operator and showing that it is bounded.

Here we consider another possibility, a classical exotic matter which, through Einstein equations, gives a nonsingular universe which behaves as the traditional models as time flows. For simplicity, we consider the flat FLRW universe. We then show that in the classical limit, the matter content which fills the universe tends to become usual matter with EoS $p=w\rho$. This indicates that it is possible to find bouncing universes and still recover the main features of usual classical cosmology as time flows. We consider two types of matter: a perfect fluid, for which we find an implicit EoS; and two non-interacting scalar fields, one of them of the phantom type, for which we will be looking for their potential energy. 

Nonlinear EoS have already been considered in previous works. For instance, in \cite{nojiri2,ananda} the general quadratic EoS $p=p_0+\alpha\rho+\beta\rho^2$ was studied and it was shown that with the right choice of the parameters $p_0$, $\alpha$ and $\beta$ this exotic fluid can model dark matter. In \cite{nojiri2}, it was shown that such EoS fits a realistic model where a decelerating universe turns into an accelerating one. More general EoS were studied in \cite{capozziello,nojiri2,brevik} and again it was demonstrated that early and late-time cosmic accelerations can be modeled by such exotic fluids. However, there are cases where we cannot find an explicit EoS of the form $p=p(\rho)$. In these cases we have an implicit EoS of the form $f(\rho,p)=0$. In Ref. \cite{stef}, motivated by an universe filled with two independent cosmological fluids with EoS $p=\alpha \rho$ and $p=\eta \rho$, \v Stefan\v ci\'c found an implicit EoS representing a single fluid, which has the same effect of those two non-interacting fluids. Another example of implicit EoS can be found in \cite{nojiri}, where, by considering an implicit EoS depending also on the Hubble parameter $H=\dot{a}/a$, it was shown that the phantom divide $w=-1$ is crossed due to the effect of the inhomogeneous term $H$ in the EoS.


A universe filled with multiple (canonical and/or phantom) scalar fields was studied in \cite{elizalde}. There, they reconstructed the entire evolution of the universe, including early inflation and late-time acceleration, using these fields. It is also possible to reconstruct the evolution of the universe through modified theories, for example $f(R)$ gravity. This can be seen in \cite{nojiri3}.

The goal of this paper is to show that, if we allow the presence of exotic matter in the universe, we can find a nonsingular universe which, as time flows, behaves as the traditional universe+perfect fluid models. However, the difference between the exotic and the ordinary matter depends on how close the initial  quantum universe is from the big bang singularity. The paper is organized as follows, in Sec. II we give a brief review of quantum cosmology. In Sec. III we find an implicit EoS of a single fluid which satisfies our requirements and show that this single fluid is equivalent to two non-interacting fluids, one being stiff matter with negative energy density. In Sec. IV we show that we can have a bouncing universe also by the introduction of two independent scalar fiels, one of the phantom type. In particular we find the potential energy of these fields. Finally, in Sec. V, we discuss the main results presented in this work.

\section{Quantum cosmology}

The action for general relativity with a perfect fluid in Schutz's formalism \cite{schutz1,schutz2} is given by (in units where $16\pi G=1$)
\begin{equation}
S_{G}=\int_{\mathcal{M}}{d^{4}x\sqrt{-g}R}+2\int_{\partial \mathcal{M}}{d^3x\sqrt{h}h_{ab}K^{ab}}+\int_{\mathcal{M}}{d^4x\sqrt{-g}p},
\label{action for general relativity2}
\end{equation}
where $h_{ab}$ is the induced metric over the boundary $\partial \mathcal{M}$ of the four-dimensional manifold $\mathcal{M}$, $K^{ab}$ is the second fundamental form of the hypersurface $\partial \mathcal{M}$ and $p$ is the  fluid pressure, which is linked to the energy density by the EoS  $p=w \rho$. Here we consider $-1\leq w <1$. Note that the stiff matter $w=1$ case is not considered here (see \cite{oliveira}). 

The super-Hamiltonian for this action in a flat FLRW universe given by the metric
\begin{equation}
ds^2=-N^2(t)dt^2+a^2(t)(dx^2+dy^2+dz^2)
\end{equation}
is given by \cite{lapichinskii}
\begin{equation}
\mathcal{H}=-\frac{p_a^2}{24 a}+\frac{p_T}{a^{3w}}\approx 0,
\label{super-Hamiltonian2}
\end{equation}
where $p_a=-12\frac{da}{dT}a/N$ is the momentum conjugated to the scale factor $a(t)$ and $p_T$ is the momentum conjugated to the dynamical degree of freedom of the fluid \cite{lapichinskii}. In Schutz's formalism \cite{schutz1,schutz2}, the action representing the fluid contribution to gravity depends on six potentials (the specific entropy and enthalpy, two potentials related with rotation of the fluid and two potentials with no clear physical meaning). However, as Lapichinskii and Rubakov showed \cite{lapichinskii}, in FRW models the degrees of freedom of the fluid are reduced to one, due to the symmetry of the universe. Therefore, the action depends on only one degree of freedom of the fluid (the dynamical variable $T$). The physical meaning of this dynamical variable is not clear, but since its conjugated momentum appears linearly in Eq. (\ref{super-Hamiltonian2}), it may work as a measure of time after a quantization process.

The super-Hamiltonian (\ref{super-Hamiltonian2}) is a constraint. Therefore, following Dirac's algorithm \cite{dirac} for quantization of constrained Hamiltonian systems, we make the substitutions
\begin{equation}
p_a\to-i\frac{\partial}{\partial a};\;\;\;\;\;p_T\to-i\frac{\partial}{\partial T}
\label{canonical quantization2}
\end{equation}
and demand that the super-Hamiltonian operator  annihilate the wave function of the universe. The result is
\begin{equation}
\frac{\partial^2\Psi}{\partial a^2}+24ia^{1-3w}\frac{\partial \Psi}{\partial t}=0,
\label{wheeler-dewitt equation2}
\end{equation}
where $t=-T$ is the time coordinate in the gauge $N(t)=a^{3w}(t)$ \cite{alvlemos}. Note that we have defined the new variable $t$ in order to have a Schr\"odinger type equation $\hat{H}\Psi=i\partial\Psi/\partial t$. This is the so called Wheeler-DeWitt equation of the universe \cite{dewitt}. 

The scale factor is defined on the half-line $(0,\infty)$, therefore with the scalar product
\begin{equation}
\left<\Psi|\Phi\right>=\int_{0}^{\infty}{a^{1-3w}\Psi(a,t)^{\ast}\Phi(a,t)da},
\end{equation}
a boundary condition in $a=0$ is necessary in order to ensure self-adjointness of Eq. (\ref{wheeler-dewitt equation2}). Lemos found in \cite{lemos} that all possible boundary conditions are given by
\begin{equation}
\Psi'(0,t)=\beta \Psi(0,t),\;\;\;\beta\in\mathbb{R},
\label{boundary conditions2}
\end{equation}
where the prime denotes the derivative with respect to the variable $a$.

For the sake of simplicity, let us consider only two possibilities $\beta=0$ and $\beta=\infty$, corresponding to Neumann and Dirichlet boundary conditions, respectively. Eq. (\ref{wheeler-dewitt equation2}) can be easily solved by separating variables \cite{alvarenga}. The resultant wave packet if we choose Neumann boundary condition can be found in Ref. \cite{pitelli} and is given by
\begin{equation}
\Psi(a,t)=\frac{\left[\frac{4}{(1-w)}\sqrt{\frac{2}{3}}\right]^{\frac{1}{3(1-w)}}}{\left[2(\gamma+it)\right]^{\frac{1+3(1-w)}{3(1-w)}}}a\exp{\left[-\frac{8}{3}\frac{a^{3(1-w)}}{(1-w)^2(\gamma+it)}\right]},
\label{wave packet}
\end{equation}
where $\gamma$ is related with the width of the initial wave packet $\Psi(a,0)$. If the initial wave packet is highly concentrated close to the singularity $a=0$, then $\gamma<<1$. If the initial wave packet is spread, we have $\gamma>>1$.

Note that taking the limit $t\to0$, where we expect to note quantum corrections, we have
\begin{equation}
\Psi(a,0)=\frac{\left[\frac{4}{(1-w)}\sqrt{\frac{2}{3}}\right]^{\frac{1}{3(1-w)}}}{\left(2\gamma\right)^{\frac{1+3(1-w)}{3(1-w)}}}a\exp{\left[-\frac{8}{3}\frac{a^{3(1-w)}}{\gamma(1-w)^2}\right]}.
\label{wp0}
\end{equation}
If $\gamma<<1$, the wave packet (\ref{wp0}) is highly concentrated around $a=0$. In this case we expect that the quantum correction on the EoS of the fluid should be appreciable in order to avoid the classical singularity, since we are too close to a classical universe with an initial big bang singularity. We will see that a great amount of matter with negative energy density has to be added in the universe in order to achieve this. However, if $\gamma>>1$, the ordinary classical matter prevails, since less effort from quantum mechanics should be put in order to remove the singularity, because the wave packet (\ref{wp0}) is spread and its peak is far away from the big bang singularity.


Given the wave packet (\ref{wave packet}), the expected value of the scale factor can be readily computed from the wave function $\Psi(a,t)$ through the formula
\begin{equation}
\left<a\right>(t)=\frac{\left<\Psi\left|a\right|\Psi\right>}{\left<\Psi|\Psi\right>},
\end{equation}
and the Bohmian trajectories \cite{holland} $a(t)$ of the scale factor can be found through the solution of the equation
\begin{equation}
p_a=\frac{\partial S}{\partial a}.
\end{equation}

The result is \cite{alvarenga}
\begin{equation}
\left<a\right>(t)=a(t)=\left(\gamma^2+t^2\right)^{\frac{1}{3(1-w)}}.
\label{scale factor}
\end{equation}

The case with Dirichlet boundary condition is very similar, so we do not discuss it here. We also ignore the possibility $w=1$ (see \cite{oliveira}).

Note that within the limits $t\to\infty$ or $\gamma<<1$, we have $a(t)\sim t^{\frac{2}{3(1-w)}}$. In the cosmic gauge, $\tau$ is given by $a^{3w}(t)dt=d\tau$. Therefore $a(\tau)\sim \tau^{2/(3(1+w))}$ for $w\neq -1$ and $a(\tau)\sim\exp{\left(\tau/3\right)}$ for $w=-1$, as expected from the classical model \cite{wald}.


\section{Exotic Equation of state}

The four-velocity of a perfect fluid in comoving coordinates in the gauge $N(t)=a^{3w}(t)$ is given by
\begin{equation}
U_{\mu}=(a^{3w}(t),0,0,0),
\end{equation}
in such a way that the energy-momentum tensor of the fluid 
\begin{equation}
T_{\mu\nu}=(\rho+p)U_{\mu}U_{\nu}+pg_{\mu\nu}
\end{equation}
is given by
\begin{equation}
T_{\mu\nu}=\text{diag}(a^{6w}\rho,a^2p,a^2p,a^2p).
\end{equation}

Einstein field equations (in units where $16\pi G=1$) become
\begin{equation}\begin{cases}
G_{00}&=3\frac{\dot{a}^2(t)}{a^2(t)}=\frac{1}{2} a^{6w}\rho\\
G_{ii}&=a^{-6w}\left[(6w-1)\dot{a}^2(t)-2a\ddot{a}\right]=\frac{1}{2} a^2(t)p,\;\;\;i=1,2,3.
\end{cases}
\label{einsteins}
\end{equation}

Substituting Eq. (\ref{scale factor}) into the above equations leads to
\begin{subequations}
\begin{align}
\rho&=\frac{8t^2}{3(1-w)^2}\left(\gamma^2+t^2\right)^{-\frac{2}{1-w}}, \label{rho1}\\
p&=w\rho-\frac{8\gamma^2}{3(1-w)}\left(\gamma^2+t^2\right)^{-\frac{2}{1-w}}.\label{p1}
\end{align}
\end{subequations}

Let us now manipulate Eqs. (\ref{rho1}) and (\ref{p1}) in order to find an implicit relation between $\rho$ and $p$. First note that
\begin{equation}
\rho-\frac{1}{1-w}p=-\frac{w}{1-w}\rho+\frac{8}{3(1-w)^2}\left(\gamma^2+t^2\right)^{-\frac{2}{1-w}+1}.
\end{equation}
Therefore
\begin{equation}\begin{aligned}
\frac{\rho-p}{1-w}&=\frac{1}{\gamma^2(1-w)}\left[\frac{8\gamma^2}{3(1-w)}\left(\gamma^2+t^2\right)^{-\frac{1+w}{1-w}}\right]\\
&=\frac{1}{\gamma^2(1-w)}\left[\frac{8\gamma^2}{3(1-w)}\left(\gamma^2+t^2\right)^{-\frac{2}{1-w}}\right]^{\frac{1+w}{2}}\\&\times \left[\frac{8\gamma^2}{3(1-w)}\right]^{-\frac{1+w}{2}+1}\\
&=\frac{1}{\gamma^2(1-w)}\left(w\rho-p\right)^{\frac{1+w}{2}}\left[\frac{8\gamma^2}{3(1-w)}\right]^{\frac{1-w}{2}}
\end{aligned}\end{equation}
Squaring both sides of the last equation and rearranging the terms we finally obtain
\begin{equation}
\left(w\rho-p\right)^{1+w}=\gamma^4\left[\frac{3(1-w)}{8\gamma^2}\right]^{1-w}(\rho-p)^2.
\end{equation}

Therefore, we have an implicit EoS given by $f(\rho,p)=0$, where
\begin{equation}
f(\rho,p)=\left(w\rho-p\right)^{1+w}-\gamma^{4}\left[\frac{3(1-w)}{8\gamma^2}\right]^{1-w}(\rho-p)^{2}.
\label{EoS}
\end{equation}

Let us show that $p=w\rho$ in the limit $t\to \infty$. For this, we use Eqs. (\ref{rho1}) and (\ref{p1}). We have
\begin{equation}
\frac{p}{\rho}=w-\frac{(1-w)}{t^2}\to w.
\end{equation} 
Therefore, $p\to w\rho$ in the ``classical'' limit. This is not surprising, since we started with an expression for the scale factor which resembles the classical behavior as $t\to\infty$.

By Eq. (\ref{p1}) we see that the first effect from quantum mechanics to the EoS of the fluid is to introduce a constant term
\begin{equation}
p(t=0)=-\frac{8}{3(1-w)}\gamma^{-2\frac{1+w}{1-w}}.
\end{equation}
This term is necessary for the initial cosmic acceleration, which prevents the singularity. This can be seen by taking $\rho(t\sim 0)=0$ and $p(t\sim 0)=-\frac{8}{3(1-w)}\gamma^{-2\frac{1+w}{1-w}}$ in Eq. (\ref{einsteins}). We then have
\begin{equation}
a''(t\sim 0)=\frac{2}{3}\gamma^{-2\frac{1+w}{1-w}}a(t)^{1+6w}>0.
\end{equation}

We can also see in Eqs. (\ref{rho1}) and (\ref{p1}) that $t=0$ corresponds to $\rho=0$ and $p=-\frac{8}{3}\gamma^{-2\frac{1+w}{1-w}}$. To analyze the behavior of the fluid in this case we expand Eq. (\ref{EoS}) in Taylor series around the point $(\rho,p)=(0,-\frac{8}{3(1-w)}\gamma^{-2\frac{1+w}{1-w}})$ and solve $p$ as a function of $\rho$. The result is 
\begin{equation}
p\approx-\frac{8}{3(1-w)}\gamma^{-2\frac{1+w}{1-w}}+(w+2)\rho.
\label{strange}
\end{equation}
Note that for $\gamma<<1$ there is a huge initial negative pressure. This is necessary for the initial cosmic acceleration necessary to escape the initial big bang singularity. For $\gamma>>1$ we have an equation of the form $p=\text{constant}\times\text{density}$. The addition of matter with negative energy changes completely the configuration of the content of the universe for small $t$. However the quantum effects are more dramatic as $\gamma<<1$.

Eq. (\ref{strange}) may seen strange, since the speed of which perturbations propagate in the fluid is given by
\begin{equation}
c_s^2=\frac{d p}{d \rho}=w+2>1.
\end{equation}
However, as we will see in a moment, this is an effective equation which corresponds to two non-interacting fluids, one of them representing stiff matter with negative energy density [or equivalently two interacting fluids satisfying the strong energy condition (see \cite{pinto-neto})]. 


For this, consider two non-interacting fluids with EoS $p=\alpha \rho$ and $p=\eta \rho$ as in Ref. \cite{stef}. The density as a function of the scale factor is given by
\begin{equation}
\rho=C_1a^{-3(1+\alpha)}+C_2a^{-3(1+\eta)}.
\end{equation}
The equation of the energy-momentum tensor conservation
\begin{equation}
d\rho+3(\rho+p)\frac{da}{a}=0
\end{equation}
implies the following equation for the pressure
\begin{equation}
p=\alpha C_1a^{-3(1+\alpha)}+\eta C_2a^{-3(1+\eta)}.
\end{equation}

Note that
\begin{equation}
\left(\frac{\alpha\rho-p}{(\alpha-\eta)C_2}\right)^{1/(1+\eta)}=a^{-3}=\left(\frac{p-\eta \rho}{(\alpha-\eta)C_1}\right)^{1/(1+\alpha)},
\end{equation}
so that if we take $\alpha=w$ and $\eta=1$ we arrive with
\begin{equation}
(w\rho-p)^{1+w}=\frac{\left[(w-1)C_2\right]^{1+w}}{(1-w)^2C_1^2}(\rho-p)^2.
\end{equation}
If we choose, for example,
\begin{equation}
C_1=\frac{1}{\gamma^2(1-w)};\,C_2=-\frac{1}{1-w}\left[\frac{3(1-w)}{8\gamma^2}\right]^{\frac{1-w}{1+w}},
\end{equation}
we recover Eq. (\ref{EoS}). Therefore, one single fluid with EoS (\ref{EoS}) is equivalent to two non-interacting fluids with linear EoS, one of them representing stiff matter with negative energy density
\begin{equation}
\rho=\frac{1}{\gamma^2(1-w)}a^{-3(1+w)}-\frac{1}{1-w}\left[\frac{3(1-w)}{8\gamma^2}\right]^{\frac{1-w}{1+w}}a^{-6}.
\label{density}
\end{equation}
Note that this procedure works for $w\neq-1$. As we will see later, the case $w=-1$ will be represented by a single fluid with EoS $p=\text{constant}+\rho$. These results match the one found in Ref. \cite{pinto-neto}, where Pinto-Neto et al. study cosmology with two interacting fluids satisfying the strong energy condition and with EOS $p=\text{contant}\times\text{density}$ which behave effectively as independent fluids, one having negative energy density.

Let us now see a few examples of the most important cases $w=1/3$, $w=0$ and $w=-1$. 

\subsection{$w=1/3$}

\begin{figure}[!htb]
\centering
\includegraphics[scale=0.5]{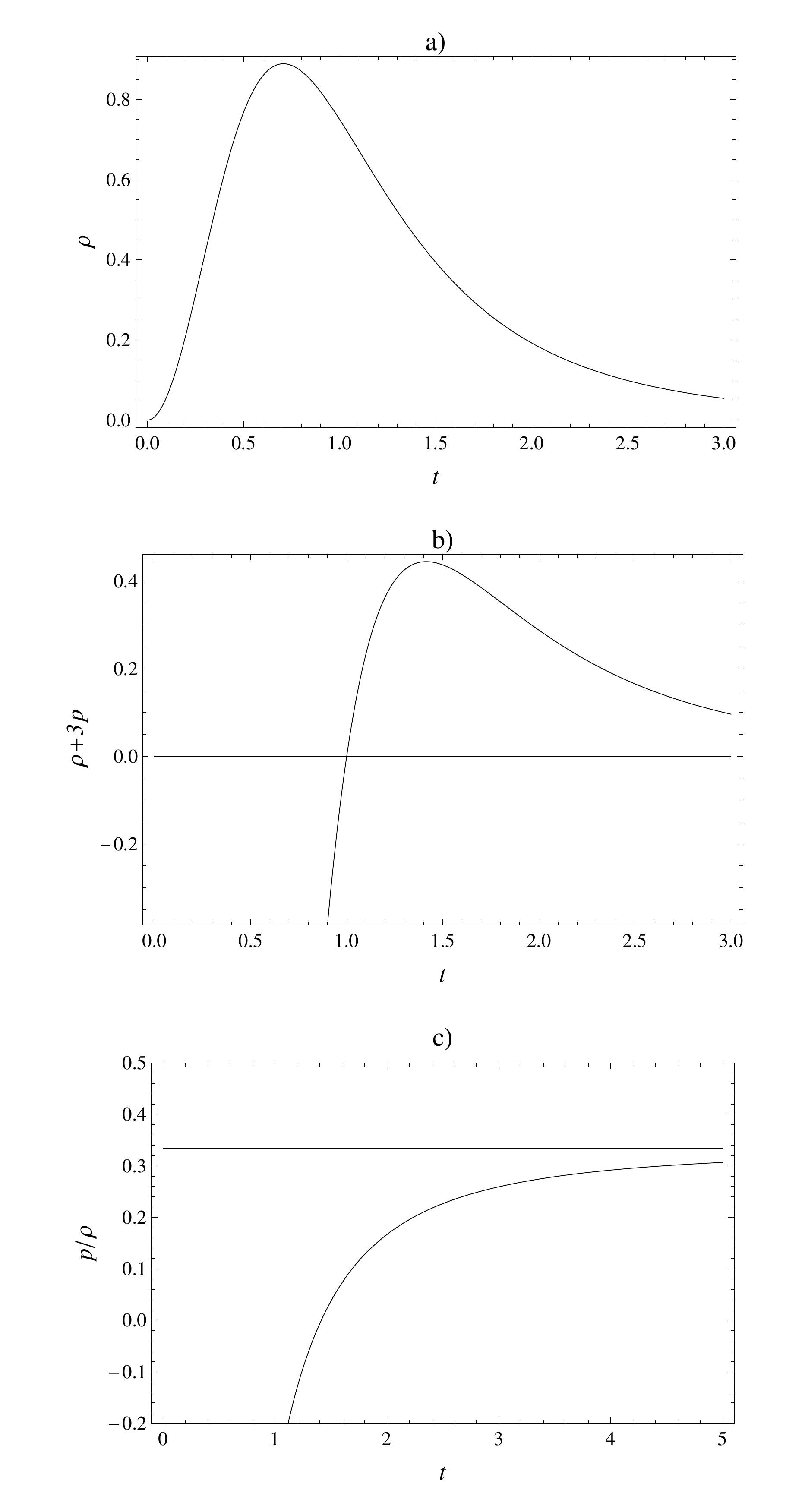}
\caption{$\left.a\right)$ Letting $\gamma=1$, we see that the curve of $\rho(t)$ starts at $0$, it has a maximum $\rho_{\text{max}}=9/8$ at $t=1/\sqrt{2}$ and it decays to $0$ as $t\to\infty$. $\left.b\right)$ Also, looking to the graphic of  $\rho(t)+3p(t)$ for the same value of $\gamma$, we clearly see that the strong energy condition is violated for small values of $t$. $\left.c\right)$. Finally, we consider the graphic of $p(t)/\rho(t)$ for $\gamma=1$. In the classical limit ($t\to\infty$), $p/\rho\to 1/3$.}
\label{fig1}
\end{figure}

For $w=1/3$, we can see in Fig. \ref{fig1}-$\left.a\right)$ that $\rho(t)$ is not a one-to-one function of time. We have to invert $t$ as a function of $\rho$ in two branches and for each branch we have a particular EoS. In this case, it is impossible for a single fluid to have a unique EoS $p=p(\rho)$ and an implicit EoS is necessary. Fig. \ref{fig1}-$\left.b\right)$ shows that the strong energy condition which states that $\rho+3p\geq0$ is violated for small $t$. This explains why initial singularity is excluded in this model. Besides, Fig. \ref{fig1}-$\left.c\right)$ shows graphically that in the ``classical'' limit, i.e., for $t\to\infty$, we have the asymptotic behavior $p\to\frac{\rho}{3}$.

\begin{figure}[!htb]
\centering
\includegraphics[scale=0.5]{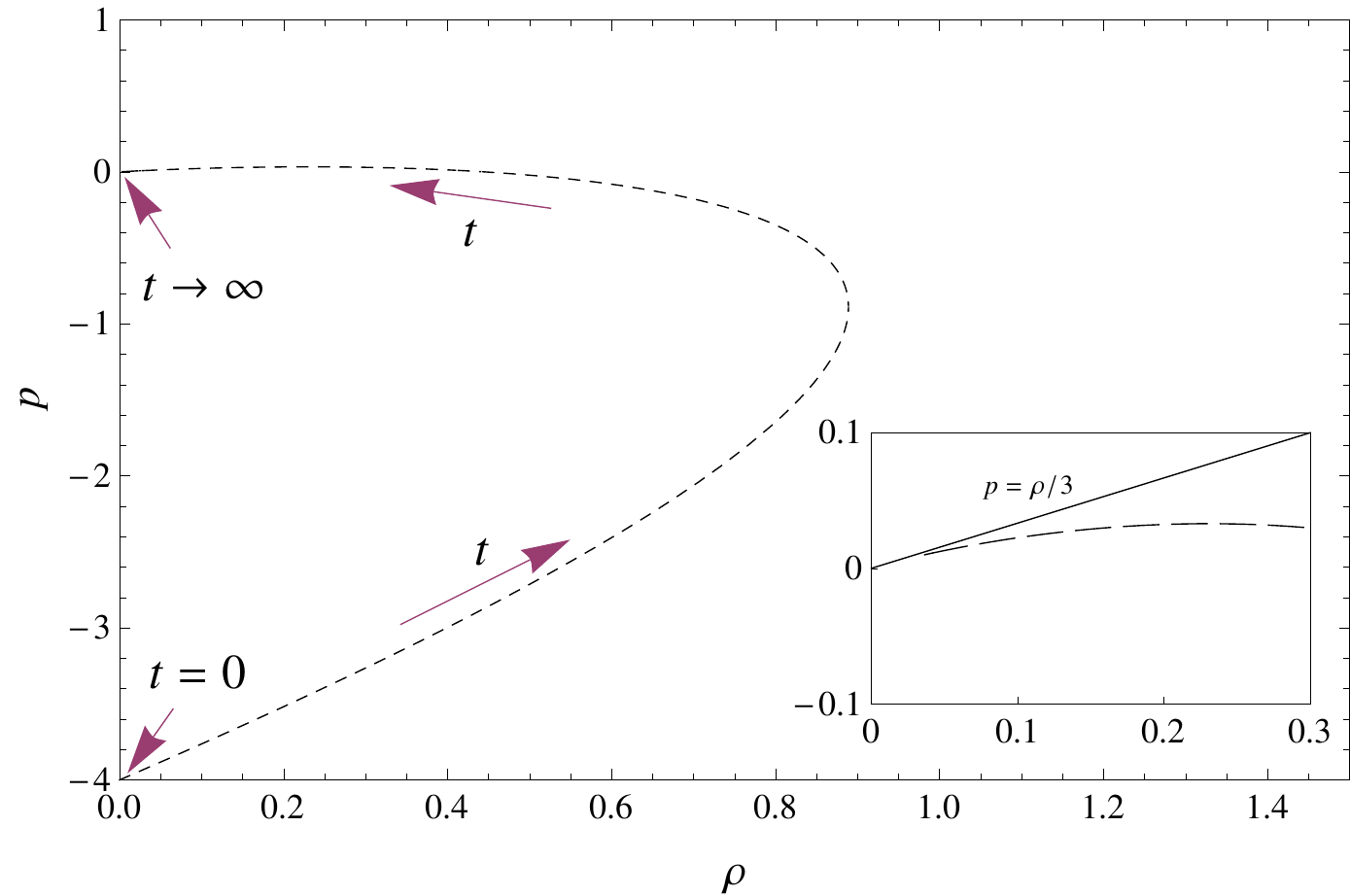}
\caption{The graph of Eq. (\ref{f}). It represents the physical trajectory of $p$ as a function of $\rho$ as time flows (see the arrows representing the direction of time). Inset we see that as $t\to\infty$, we recover the classical EoS $p=\rho/3$.}
\label{newfig}
\end{figure}

Returning to Eq. (\ref{EoS}) we have
\begin{equation}
f_{\text{radiation}}(\rho,p)=\left(\frac{1}{3}\rho-p\right)^{4/3}-\gamma^4\left(\frac{1}{4\gamma^2}\right)^{2/3}(\rho-p)^2=0,
\end{equation}
or equivalently
\begin{equation}
\left(\frac{1}{3}\rho-p\right)^2-\frac{\gamma^4}{4}(\rho-p)^3=0.
\label{f}
\end{equation}
The curve representing the physical trajectory of $p$ as a function of $\rho$ is displaced in Fig. \ref{newfig}.

We can use Eq. (\ref{density}) to analyse which fluid prevails in the quantum domain ($t<<1$) according to the size of $\gamma$. Substituting $w=1/3$ and $a(t)=(\gamma^2+t^2)^{1/2}$ in Eq. (\ref{density}) leads to
\begin{equation}
\rho(t)=\frac{3}{2\gamma^2}(\gamma^2+t^2)^{-2}-\frac{3}{4\gamma}(\gamma^2+t^2)^{-3}.
\end{equation}
By expanding the above equation in Taylor series around $t=0$ we have
\begin{equation}
\rho(t)\approx -\frac{3}{4\gamma^7}\left[1-3\left(\frac{t}{\gamma}\right)^2\right]+\frac{3}{2\gamma^6}\left[1-2\left(\frac{t}{\gamma}\right)^2\right].
\end{equation}
We conclude that for $\gamma<<1$, most of the contribution to the effective equation (\ref{EoS}) comes from the stiff matter with negative energy density. If $\gamma>>1$, the ordinary fluid $p=\rho/3$ prevails. The quantum mechanics action of introducing an effective fluid with negative energy density is greater when $\gamma$ is small. As we said before, this is the limit of a classical universe, so  we expect a greater effect. Clearly, the effective introduction of a fluid with negative energy density is necessary for all values of $\gamma$. However, its significance depends on which limit the initial universe is characterized.

\subsection{$w=0$}
For $w=0$, Eq. (\ref{EoS}) has the form
\begin{equation}
f_{\text{dust}}(\rho,p)=p+\frac{3\gamma^2}{8}(\rho-p)^2=0.
\end{equation}
A simple inspection in Eq. (\ref{p1}) shows that in this case $p\leq0$ so that the above equation makes sense. If we solve this equation for $p$ as a function of $\rho$ we find two branches:
\begin{equation}\begin{aligned}
\text{branch I:}&\,\,\,p=-\frac{4}{3\gamma^2}-\frac{1}{3\gamma^2}\sqrt{16-24\gamma^2\rho}+\rho,\\
\text{branch II:}&\,\,\,p=-\frac{4}{3\gamma^2}+\frac{1}{3\gamma^2}\sqrt{16-24\gamma^2\rho}+\rho.
\end{aligned}\end{equation}
Note the presence of a stiff matter component in the expressions for $p(\rho)$. 

\subsection{$w=-1$}

For $w=-1$, the classical universe is non-singular and the expansion is exponential. Quantum mechanics is not necessary to avoid any singularity, so it works in the same way for any value of $\gamma$. The quantum universe can be reproduced by a single fluid with the pressure having a negative constant term plus a stiff matter contribution. As we will see this type of fluid causes inflation as well as dark matter. 

The expression for $\rho$ as a function of $t$ is given by
\begin{equation}
\rho(t)=\frac{2t^2}{3}\frac{1}{\gamma^2+t^2}.
\end{equation}
Clearly, this is a one-to-one function, so we can find a simple EoS $p=p(\rho)$. In fact, Eq. (\ref{EoS}) becomes in this case
\begin{equation}
f_{\text{DE}}(\rho,p)=1-\frac{9}{16}(\rho-p)^2=0.
\end{equation}
We can solve this equation for $p$ as a function of $\rho$. It is easy to find
\begin{equation}
p=\pm\frac{4}{3}+\rho.
\end{equation}
Looking at Eq. (\ref{p1}) we see that for $\rho=0$ we have $p\leq 0$, so that the right expression for the EoS in this case is given by
\begin{equation}
p=-\frac{4}{3}+\rho.
\label{w-1}
\end{equation}
Note again the presence of a stiff matter term. This kind of EoS was already studied in Ref. \cite{babichev}. There, they studied models of fluids with EoS of the form $p=\alpha(\rho-\rho_0)$, with $\alpha\geq 0$, and showed that it is possible to include dark (and also phantom) energy with a positive squared sound speed of linear pertubations $c_s^2\equiv \partial p/\partial \rho=\alpha\geq 0$. With this EoS in hands we can solve Einstein equations in the cosmic gauge. The $\mu\nu=ij$ equation is given by (in units where $16\pi G=1$)
\begin{equation}
\frac{\ddot{a}}{a}+2\left(\frac{\dot{a}}{a}\right)^2=\frac{1}{4}(\rho-p)=\frac{1}{3},
\end{equation}
where $\dot{a}=da/d\tau$ and $\tau$ is the cosmic time.
This equation, with the initial condition $a(0)=\gamma^{1/3}$ [see Eq. (\ref{scale factor})], can be solve and the result is
\begin{equation}
a(\tau)=\gamma^{1/3}\cosh^{1/3}(\tau).
\end{equation}
We find the same expression for $a(\tau)$ if we integrate the equation $a^{-3}(t)dt=d\tau$, which relates two measures of time in different gauges. Note that the constant term in Eq. (\ref{w-1}) works as a cosmological constant.

\section{Scalar fields}

We can also try to reproduce the results of the last section with the use of two scalar fields instead of a perfect fluid. We choose to add a phantom and a standard scalar field. As we will see in a moment, with these two non-interacting fields we can fit the initial cosmic acceleration and still recover the classical behavior of the universe as $t\to\infty$. The process of reconstruction of the of the universe in this case is analogous to the previous case, so that the fields act effectively as a perfect fluid with EoS give by Eq. (\ref{EoS}). 

The Lagrangian for these two fields is given by
\begin{equation}\begin{aligned}
L=&-\frac{1}{2}g^{\mu\nu}\partial_\mu\phi_{\text{st}}\partial_\nu\phi_{\text{st}}-V(\phi_{\text{st}})\\&+\frac{1}{2}g^{\mu\nu}\partial_\mu\phi_{\text{ph}}\partial_\nu\phi_{\text{ph}}-U(\phi_{\text{ph}}),
\end{aligned}\end{equation}
where $\phi_{\text{st}}$ corresponds to the standard scalar field while $\phi_{\text{ph}}$ corresponds to the phantom one.

The energy-momentum tensor $T_{\mu\nu}$ in the standard case is given by
\begin{equation}
T_{\mu\nu}^{\text{st}}=\partial_{\mu}\phi_{\text{st}}\partial_{\nu}\phi_{\text{st}}+g_{\mu\nu}\left[-\frac{1}{2}g^{\lambda\sigma}\partial_{\lambda}\phi_{\text{st}}\partial_{\sigma}\phi_{\text{st}}-V(\phi_{\text{st}})\right],
\end{equation} 
and in the phantom case is given by
\begin{equation}
T_{\mu\nu}^{\text{ph}}=-\partial_{\mu}\phi_{\text{ph}}\partial_{\nu}\phi_{\text{ph}}+g_{\mu\nu}\left[\frac{1}{2}g^{\lambda\sigma}\partial_{\lambda}\phi_{\text{ph}}\partial_{\sigma}\phi_{\text{ph}}-U(\phi_{\text{ph}})\right].
\end{equation}
Therefore, the  components of the total energy-momentum tensor $T_{\mu\nu}=T_{\mu\nu}^{\text{st}}+T_{\mu\nu}^{\text{ph}}$ in the gauge $N(t)=a^{3w}(t)$ are given by
\begin{equation}\begin{aligned}
T_{00}&=\frac{\dot{\phi}_{\text{st}}^2}{2}-\frac{\dot{\phi}_{\text{ph}}^2}{2}+a^{6w}V(\phi_{\text{st}})+a^{6w}U(\phi_{\text{ph}})=a^{6w}\rho,\\
T_{ii}&=\frac{\dot{\phi}_{\text{st}}^2}{2a^{6w-2}}-\frac{\dot{\phi}_{\text{ph}}^2}{2a^{6w-2}}-a^2V(\phi_{\text{st}})-a^2V(\phi_{\text{ph}})=a^2p.
\end{aligned}\end{equation}
Therefore, using Eqs. (\ref{rho1}) and (\ref{p1}) we have
\begin{equation}\begin{aligned}
\rho+p&=\frac{\dot{\phi}_{\text{st}}^2}{a^{6w}}-\frac{\dot{\phi}_{\text{ph}}^2}{a^{6w}}\\&=\frac{8}{3(1-w)^2}\left[\frac{t^2(1+w)-(1-w)\gamma^2}{(\gamma^2+t^2)^{\frac{2}{1-w}}}\right],\\
\rho-p&=2V(\phi_{\text{st}})+2U(\phi_{\text{ph}})=\frac{8}{3}\left[\frac{1}{(\gamma^2+t^2)^{\frac{1+w}{1-w}}}\right].
\end{aligned}\end{equation}

Using Eq. (\ref{scale factor}) we have
\begin{subequations}
\begin{align}
&\dot{\phi}_{\text{st}}^2-\dot{\phi}_{\text{ph}}^2=\frac{8}{3(1-w)^2}\left[\frac{t^2(1+w)-(1-w)\gamma^2}{(\gamma^2+t^2)^2}\right],\label{phiphi}\\
&V(\phi_{\text{st}})+U(\phi_{\text{ph}})=\frac{4}{3(1-w)}\left[\frac{1}{(\gamma^2+t^2)^{\frac{1+w}{1-w}}}\right].\label{VV}
\end{align}
\label{51}\end{subequations}

There are a positive and a negative contributions on both sides of Eq. (\ref{phiphi}). One possible way to fulfill Eq. (\ref{phiphi}) is given by
\begin{equation}\begin{aligned}
\dot{\phi}_{\text{st}}&=\frac{2\sqrt{2}}{\sqrt{3}}\frac{t\sqrt{1+w}}{(1-w)(\gamma^2+t^2)},\\
\dot{\phi}_{\text{ph}}&=\frac{2\sqrt{2}}{\sqrt{3(1-w)}}\frac{\gamma}{(\gamma^2+t^2)}.
\end{aligned}\end{equation} 
Note that, in this paper, we are searching for a classical model which gives a bouncing universe which behaves as the traditional models for later times. As we will see, by choosing the form of $\dot{\phi}_{\text{st}}$ and $\dot{\phi}_{\text{ph}}$  above we will find two particular potentials for the fields. However, there are other ways to fulfill Eq. (\ref{phiphi}), which would lead to other forms for the energy potentials.

Integrating the above equations leads to
\begin{equation}\begin{aligned}
\phi_{\text{st}}&=\phi_{0\text{st}}+\frac{\sqrt{2}}{\sqrt{3}}\frac{\sqrt{1+w}}{1-w}\ln{\left(1+t^2/\gamma^2\right)},\\
\phi_{\text{ph}}&=\phi_{0\text{ph}}+\frac{2\sqrt{2}}{\sqrt{3(1-w)}}\arctan{\left(t/\gamma\right)}.
\end{aligned}\label{phidet}\end{equation}

Because 
\begin{equation}
\dot{\phi}_{\text{st}}(0)=0;\,\,\,\dot{\phi}_{\text{ph}}(0)=\frac{2\sqrt{2}}{\gamma\sqrt{3(1-w)}},
\end{equation}
the contribution from the phantom field to the kinetic energy prevails for $t<<1$. This is necessary for initial cosmic acceleration. As time increases, the contribution from the standard scalar field dominates. Note that the period when the phantom field prevails depends on $\gamma$. If $\gamma$ is small (initial universe close to the big bang), the phantom field prevais for a longer time. If $\gamma$ is large, the phantom field is necessary only for a short period of time.

We can use Eq. (\ref{phidet}) to invert $t$ as a function of the fields $\phi_{\text{st}}$ and $\phi_{\text{ph}}$ in two different ways. The result is
\begin{subequations}\begin{align}
t^2+\gamma^2&=\gamma^2\exp{\left[\sqrt{\frac{3}{2(1+w)}}(1-w)\left(\phi_{\text{st}}-\phi_{0\text{st}}\right)\right]},\label{t1}\\
t&=\gamma\tan{\left[\frac{\sqrt{3(1-w)}}{2\sqrt{2}}\left(\phi_{\text{ph}}-\phi_{0\text{ph}}\right)\right]}\label{t2}.
\end{align}\end{subequations}
If we substitute Eq. (\ref{t1}) in Eq. (\ref{VV}) we find a possible form for the sum of the potential energies
\begin{equation}
V(\phi_{\text{st}})+U(\phi_{\text{ph}})=\frac{4}{3(1-w)\gamma^{\frac{2(1+w)}{1-w}}}\exp{\left[-\sqrt{\frac{3(1+w)}{2}}\left(\phi_{\text{st}}-\phi_{0\text{st}}\right)\right]}.
\end{equation}
If we use Eq. (\ref{t2}) instead, we find
\begin{equation}
V(\phi_{\text{st}})+U(\phi_{\text{ph}})=\frac{4}{3(1-w)\gamma^{\frac{2(1+w)}{1-w}}}\frac{1}{\sec^{\frac{2(1+w)}{1-w}}{\left[\frac{\sqrt{3(1+w)}}{2\sqrt{2}}\left(\phi_{\text{ph}}-\phi_{0\text{ph}}\right)\right]}}.
\end{equation}

A possible way to satisfy the last two equations comes with the introduction of a constant parameter $0\leq\alpha\leq1$ in the following way
\begin{equation}\begin{aligned}
V(\phi_{\text{st}})&=\alpha\frac{4}{3(1-w)\gamma^{\frac{2(1+w)}{1-w}}}\exp{\left[-\sqrt{\frac{3(1+w)}{2}}\left(\phi_{\text{st}}-\phi_{0\text{st}}\right)\right]},\\
U(\phi_{\text{ph}})&=(1-\alpha)\frac{4}{3(1-w)\gamma^{\frac{2(1+w)}{1-w}}}\times\frac{1}{\sec^{\frac{2(1+w)}{1-w}}{\left[\frac{\sqrt{3(1-w)}}{2\sqrt{2}}\left(\phi_{\text{ph}}-\phi_{0\text{ph}}\right)\right]}}.
\end{aligned}\end{equation}

\subsection{$w=1/3$}

In this case we have the following expressions for the fields as functions of $t$
\begin{equation}\begin{aligned}
\phi_{\text{st}}(t)&=\phi_{0\text{st}}+\sqrt{2}\ln(1+t^2/\gamma^2),\\
\phi_{\text{ph}}(t)&=\phi_{0\text{ph}}+2\arctan{t/\gamma}.
\end{aligned}\end{equation}

On the other hand, the expressions for the potential energies are given by
\begin{equation}\begin{aligned}
V(\phi_{\text{st}})&=\alpha\frac{2}{\gamma^4}\exp{\left[-\sqrt{2}(\phi_{\text{st}}-\phi_{0\text{st}})\right]},\\
U(\phi_{\text{ph}})&=(1-\alpha)\frac{2}{\gamma^4\left[1+\tan^2{\left(\frac{\phi_{\text{ph}}-\phi_{0\text{ph}}}{2}\right)}\right]^2}.
\end{aligned}\end{equation}

If we define the quantity $w(t)$ by
\begin{equation}
w(t)=\frac{p}{\rho}=\frac{\frac{\dot{\phi}_{\text{st}}^2}{2a^{6w}}-\frac{\dot{\phi}_{\text{ph}}^2}{2a^{6w}}-V(\phi_{\text{st}})-V(\phi_{\text{ph}})}{\frac{\dot{\phi}_{\text{st}}^2}{2a^{6w}}-\frac{\dot{\phi}_{\text{ph}}^2}{2a^{6w}}+V(\phi_{\text{st}})+V(\phi_{\text{ph}})},
\label{w(t)}
\end{equation}
it is easy to show that as $t\to\infty$ we have $w(t)\to1/3$. Obviously the same works for any value of $w$.

\subsection{$w=0$}

This case is not much different from the previous one. The expressions for the fields are given by
\begin{equation}\begin{aligned}
\phi_{\text{st}}(t)&=\phi_{0\text{st}}+\sqrt{\frac{2}{3}}\ln(1+t^2/\gamma^2),\\
\phi_{\text{ph}}(t)&=\phi_{0\text{ph}}+2\sqrt{\frac{2}{3}}\arctan{t/\gamma}.
\end{aligned}\end{equation}
The potential energies are given by
\begin{equation}\begin{aligned}
V(\phi_{\text{st}})&=\alpha\frac{4}{3^2\gamma^2}\exp{\left[-\sqrt{\frac{3}{2}}(\phi_{\text{st}}-\phi_{0\text{st}})\right]},\\
U(\phi_{\text{ph}})&=(1-\alpha)\frac{4}{3\gamma^2\left\{1+\tan^2{\left[\frac{1}{2}\sqrt{\frac{3}{2}}\left(\frac{\phi_{\text{ph}}-\phi_{0\text{ph}}}{2}\right)\right]}\right\}}.
\end{aligned}\end{equation}

In particular, both standard and phantom fields are necessary to reproduce the scalar factor (\ref{scale factor}).

\subsection{$w=-1$}

For $w=-1$, the phantom scalar field is alone sufficient to reconstruct the quantum scale factor. The substitution $w=-1$ in Eq. (\ref{phidet}) yields
\begin{equation}\begin{aligned}
\phi_{\text{st}}&=\phi_{0\text{st}},\\
\phi_{\text{ph}}&=\phi_{0\text{ph}}+\frac{2}{\sqrt{3}}\arctan{\left(t/\gamma\right)}.
\end{aligned}
\end{equation}

For the potential energies we have
\begin{equation}\begin{aligned}
V(\phi_{\text{st}})&=\alpha\frac{2}{3},\\
U(\phi_{\text{ph}})&=(1-\alpha)\frac{2}{3}.
\end{aligned}\end{equation}

The standard scalar field does not evolve in time. On the other hand the phantom one evolves but has a constant potential energy, so the system behaves as if a single free phantom scalar field was present.

Using Eq. (\ref{w(t)}) we can find the EoS of the fluid representing this configuration of fields. Note that we only have contribution to the kinetic energy from the phantom field. Therefore,
\begin{equation}\begin{aligned}
w(t)&=\frac{p}{\rho}=\frac{-\frac{\dot{\phi}_{\text{ph}}^2}{2a^{6w}}-V(\phi_{\text{st}})-V(\phi_{\text{ph}})}{-\frac{\dot{\phi}_{\text{ph}}^2}{2a^{6w}}+V(\phi_{\text{st}})+V(\phi_{\text{ph}})}\\&=-\frac{t^2+2\gamma^2}{t^2}.
\end{aligned}\end{equation}
Using Eq. (\ref{rho1}) we have
\begin{equation}
w=\frac{p}{\rho}=-\frac{4}{3\rho}+1.
\end{equation}
Therefore
\begin{equation}
p=-\frac{4}{3}+\rho.
\end{equation}

It is known that a free scalar field corresponds to a fluid with EoS $p=\rho$. In our case the field must have a constant potential energy. Note that this potential energy is independent of $\gamma$, since the universe is originally nonsingular.

\section{Final Remarks}

We have found an exotic classical fluid which reproduces the scale factor obtained in the minisuperspace quantization of the FRW universe. This fluid obeys an implicit EoS. This equation of state approaches $p=w\rho$ as $t\to\infty$ (as $\rho$ and $p$ go to zero with the expansion of the universe), so that we recover the usual behavior of the universe as time flows. The fluid satisfying the implicit EoS is equivalent to two non-interacting fluids, one of them representing stiff matter with negative energy density.  If we analyze the fluid filling the universe as a single fluid with an exotic EoS, we find that for small values of $t$, the EoS representing this fluid is very different from the ordinary matter $p=w\rho$. However, this difference can be explained by means of two non-interacting fluids, one of them with negative energy density. This negative energy is necessary in order to remove the classical singularity. The closer the initial universe is from the big bang singularity, the more is necessary to add matter with negative energy density. When we try to reproduce a classically nonsingular universe, we find that the exotic matter does not depend on the original wave packet, since there is no classical singularity to be avoided. We can also obtain the same features with the insertion of two non-interacting scalar fields, one of them being usual and the other of the phantom type. We are free to choose the fraction of potential energy for each field. In particular, we can set one of them to be free.  In any case, we have shown that it is possible to get important results from quantum cosmology and still recover the classical limit with the use of classical matter.

\acknowledgements 
J.P.M.P. thanks Alberto Saa for enlightening conversations. This work was supported by FAPESP. One of the authors, Patricio S. Letelier, sadly passed away at the end of this work. J.P.M.P. acknowledges his partnership in this and other papers.



\end{document}